# The low-entropy hydration shell at the binding site of spike RBD determines the contagiousness of SARS-CoV-2 variants

Lin Yang [a,b,1*], Shuai Guo [a,1], Chengyu Hou [c, 1], Jiacheng Li[a,1], Liping Shi[a], Chenchen Liao [c], Rongchun Shi[a], Xiaoliang Ma[a], Bing Zheng [d], Yi Fang[e], Lin Ye[b], Xiaodong He[a,f,*]

[a] National Key Laboratory of Science and Technology on Advanced Composites in Special Environments, Center for Composite Materials and Structures, Harbin Institute of Technology, Harbin 150080, China

[b] School of Aerospace, Mechanical and Mechatronic Engineering, The University of Sydney, NSW 2006, Australia

[c] School of Electronics and Information Engineering, Harbin Institute of Technology, Harbin 150080, China

[d] Key Laboratory of Functional Inorganic Material Chemistry (Ministry of Education) and School of Chemistry and Materials Science, Heilongjiang University, Harbin 150001, P. R. China.

[e] Mathematical Science Institute, The Australian National University, Canberra, ACT 0200, Australia.

[f] Shenzhen STRONG Advanced Materials Research Institute Co., Ltd, Shenzhen 518035, P. R. China.

## Abstract

The infectivity of SARS-CoV-2 depends on the binding affinity of the receptor-binding domain (RBD) of the spike protein with the angiotensin converting enzyme 2 (ACE2) receptor. The calculated RBD-ACE2 binding energies indicate that the difference in transmission efficiency of SARS-CoV-2 variants cannot be fully explained by electrostatic interactions, hydrogen-bond interactions, van der Waals interactions, internal energy, and nonpolar solvation energies. Here, we demonstrate that low-entropy regions of hydration shells around proteins drive hydrophobic attraction between shape-matched low-entropy regions of the hydration shells, which essentially coordinates protein-protein binding in rotational-configurational space of mutual orientations and determines the binding affinity. An innovative method was used to identify the low-entropy regions of the hydration shells of the RBDs of multiple SARS-CoV-2 variants and the ACE2. We observed integral low-entropy regions of hydration shells covering the binding sites of the RBDs and matching in shape to the low-entropy region of hydration shell at the binding site of the ACE2. The RBD-ACE2 binding is thus found to be guided by hydrophobic collapse between the shape-matched low-entropy regions of the hydration shells. A measure of the low-entropy of the hydration shells can be obtained by counting the number of hydrophilic groups expressing hydrophilicity within the binding sites. The low-entropy level of hydration shells at the binding site of a spike protein is found to be an important indicator of the contagiousness of the coronavirus.

*Corresponding author. E-mail address: linyang@hit.edu.cn (Lin Yang) Hexd@hit.edu.cn (Xiaodong He) [1]These authors contributed equally to this work.

## 1. Introduction

The contagiousness of a coronavirus arises from a specific physical binding between a spike protein of the coronavirus and an acceptor protein on a host cell, which in turn triggers phagocytosis of the host cell. In particular, the high affinity binding between the receptor-binding domain (RBD) of the spike protein of the novel severe acute respiratory syndrome coronavirus 2 (SARS-CoV-2) and the angiotensin converting enzyme 2 (ACE2) receptor results in the once-in-a-century, worldwide pandemic. Protein-protein binding is a spontaneous physical contact of high specificity established between two specific protein molecules in intracellular and extracellular fluid (*1-5*). The physical mechanism responsible for protein-protein binding can be considered the most important mechanism of viral infection. As a typical spontaneous reaction, protein-protein binding is mainly steered by hydrogen bonding forces, electrostatic forces, van der Waals forces, and hydrophobic interaction forces (*6, 7*) (*8, 9*). The RBD-ACE2 binding energies for main SARS-CoV-2 variants were calculated by considering electrostatic interaction, hydrogen-bond interaction, van der Waals interaction, internal energy and nonpolar solvation energy (See Fig.1 and the Supplementary) (*10-18*). Clearly, the calculated RBD-ACE2 binding energies do not explain the differences in transmission efficiency among the SARS-CoV-2 variants.

The water molecules within the hydration shells around proteins' hydrophobic groups should have a lower entropy value than bulk water molecules (*19-22*). Hydrophobic interactions between two proteins during their binding are physically powered by gradually removing ordered water molecules from the low-entropy hydration shells around hydrophobic groups at the binding site of the individual protein, which increases the entropy. The hydration shell (i.e. hydration layer) around a protein has been experimentally found to have dynamics distinct from the bulk water to a distance of 1nm (*19, 20*). This means that strong hydrophobic interactions between two proteins must have a starting distance greater than 2nm. The experiments also have shown that water molecules slow down tremendously when they enter a protein's hydration shell, and their speed is reduced by 99% (*19, 20*). Thus, the standard molar entropy of water within the ordered cages (i.e., hydration shells) around the nonpolar surface is approximately equal to the standard molar entropy of solid water, and that is about 41 J/mol/K. The standard molar entropy of liquid water is about 70 J/mol/K (*23*). Therefore, moving an ordered water molecule from the low-entropy hydration shell to free liquid results in a molar entropy difference $\Delta$ S of about 29 J/mol/K. At the human body temperature of T = 309K, entropy increment is T $\Delta$ S = 8961 J/mol for one removed hydration shell water molecule. The density of ordered water molecules within protein hydration shells is about 33 water molecules per 1 nm$^3$. In considering that the hydrophobic surface area of SARS-CoV-2 RBD involved in the hydrophobic interaction docking with the ACE2 is about 867.4 Å$^2$ (*24*), the hydrophobic attractive forces between the ACE2 and the RBD is about 4.19 nN at the starting distance about 2nm. The sum of hydrogen bonding force, electrostatic attraction force, van der Waals forces in-between the ACE2 and RBD at the distance of 2nm can be calculate by using the following equation.

$$F = F_e + F_{vw} + F_{hb} \approx \sum \frac{KQ_jQ_k}{r^2} + \sum \nabla_r(-\frac{A}{r^6} + \frac{B}{r^{12}}) + \sum \nabla_r\{[\frac{C}{r^6} - \frac{D}{r^4}]COS^4\theta\} \qquad (1)$$

Where $F_e$ is the electrostatic force, $F_{vw}$ is the van der Waals force, $F_{hb}$ is the hydrogen bonding force, $Q_j$, $Q_k$ is the amount of charge, $K$ is the electrostatic force constant , $\theta$ is the angle DHP where D is the protein donor atom, H is the hydrogen, and P is the probe accepting the hydrogen bond, A, B, C, D are the parameters determined by the atom, and r is the distance between charged atoms in between proteins. The sum of the hydrogen bonding force, electrostatic attraction force, and van der Waals forces between the ACE2 and RBD at a distance of 2nm is about 0.032nN, which is negligible compared with the hydrophobic interaction force. Between ACE2 and RBD in the distance range of 0.7nm-2.4nm, the hydrophobic interaction force accounts for more than 90 percent of the attractive force between the ACE2 and RBD (see Fig.2). Therefore, protein-protein binding is mainly guided by the hydrophobic interaction force between low-entropy regions of hydration shells of individual proteins at the guidance stage. Moreover, the prerequisite of electrostatic interaction, hydrogen-bond interaction, van der Waals interaction in-between two proteins is that the surface hydrophilic groups at the proteins' binding sites escaped from hydrogen bonding with the surrounding strong polar water molecules that require entropy-enthalpy compensation (*20, 25-29*). Therefore, only the hydrophobic interactions can be considered as long-range intermolecular attractions between proteins in coordinate protein-protein binding.

## 2 Results

Protein surface hydration dynamics are highly heterogeneous over the global protein surface (*30-32*), which proves the existence of the low-entropy regions of protein hydration shells (*33-37*). Protein-protein binding should start from the long-range hydrophobic attraction between low-entropy regions of protein's hydration shells (*38, 39*). Although the distribution of hydrophilic and hydrophobic groups on proteins' binding sites appears disorderly, the hydration shell regions that covering proteins' binding sites may be described as being low-entropy. It is possible that the hydrophilic groups at a proteins' binding site do not express their hydrophilicity, promoting the formation of a low-entropy hydration shell at the binding site. Hydrophilicity of a hydrophilic group is expressed via hydrogen bonding with surrounding water molecules. If a hydrophilic group can't hydrogen bond with surrounding water molecules, we can consider the hydrophilic group can't express its hydrophilicity. It is worth noting that protein intramolecular hydrogen bonds saturate many of the hydrogen bonds formed by surface hydrophilic groups of proteins, thereby preventing these hydrophilic groups from hydrogen bonding with surrounding water molecules. Namely, when an intramolecular-hydrogen-bonded hydrophilic group is located on a protein surface, its hydrophilicity is negligible. To form the intramolecular hydrogen bonds, nascent unfolded polypeptide chains need to escape from hydrogen bonding with surrounding polar water molecules that require entropy-enthalpy compensations during the protein folding, according to the Gibbs free energy equation (*40*). The entropy-enthalpy compensations are initially driven by laterally hydrophobic interactions among the side-chains of adjacent residues in the sequences of unfolded protein chains (*40*). As a result of the entropy-enthalpy compensations, surrounding water molecules cannot break the intramolecular hydrogen bonds of proteins by competing with the donors and acceptors of these intramolecular hydrogen bonds (*40*). It can be concluded that hydrophobic-group-induced low-entropy hydration shells can cover neighbored intramolecular-hydrogen-bonded hydrophilic groups on protein surface, due to the hydrophilicity of these hydrophilic groups having been expressed by the intramolecular hydrogen bonds. Therefore, protein surface hydration dynamics are not simply determined by the distribution of the hydrophilic and the hydrophobic groups on a protein surface (*30, 31*), but by the intramolecular-hydrogen-bonds free hydrophilic groups and the hydrophobic groups. It is worth noting that parallel distributed state of adjacent peptide planes promise the hydrogen bonding between the carbonyl oxygen atom and the adjacent amide hydrogen atom in peptide plane (*40*).

When a hydrophobic side-chain locate at the binding site of protein and protrude outward to surrounding water molecules, it can shield the backbone carbonyl oxygen atoms and amide hydrogen atoms of the residue from water molecules. In this way, the ordered water molecules in the low-entropy hydration shell of the hydrophobic side-chain are inhibited from fluctuating and rearrangement, and are therefore prevented from frequently hydrogen bonding with the backbone carbonyl oxygen atoms and amide hydrogen atoms, as shown in Fig.1a. This indicates that highly hydrophobic side-chains can inhibit their neighbored backbone carbonyl oxygen atoms and amide hydrogen atoms to rearrange their hydrogen bonding with different water molecules, namely, shielding the hydrophilicity of the backbone atoms. Valine, isoleucine, leucine, tryptophan, tyrosine, methionine, cysteine, lysine, phenylalanine, arginine, histidine, proline and alanine residues contain highly hydrophobic structures in their side-chains (see Fig.3). The hydration shells covering the backbone carbonyl oxygen groups and the amide hydrogen groups of these residues should be considered low-entropy state due to the shielding effect. It is worth noting that glycine have very weak hydrophilicity (*41-44*). The hydration shells surrounding the backbone carbonyl oxygen group and the amide hydrogen group of glycine should be considered low-entropy hydration shells.

According to hydrophobicity scales of amino acid residues, tryptophan and tyrosine are classified as hydrophobic residues with long side-chains (*41-44*). This because the long side-chains of tryptophan and tyrosine residues are mainly composed of hydrophobic alkyl and benzene ring structures. Neither of the two residues can substantially express hydrophilicity via the tiny CO or NH group at the end of the hydrophobic side-chain since most neighboring water molecules have been fixed in the long-hydrophobic-residues-induced ordered network (see Fig. 3) (*40*). Lysine also has a long side-chain composed of hydrophobic alkyl and one NH group. Therefore, the hydration shells surrounding tryptophan, tyrosine, and lysine can be considered low-entropy hydration shells.

Using the above analysis, we are able to identify hydrophilicity-unexpressed hydrophilic groups on protein surfaces. Low-entropy regions of the hydration shell of a given protein can be mapped by searching the hydrophilicity-unexpressed hydrophilic groups and hydrophobic groups on the protein surface (see Fig. 4). Binding sites on proteins should be typically covered by large low-entropy regions of the hydration shell to trigger the long-range intermolecular hydrophobic attractions between proteins. To prove this, we map the low-entropy regions of the hydration shells for 50 protein quaternary structures, and found out that binding sites on protein subunits are typically covered by the largest low-entropy regions of hydration shells of each subunit (see Fig.5 and supplementary 2). Protein quaternary structures are formed via the protein binding processes of the subunits. Hydrophilic groups located at proteins' binding sites normally do not express their hydrophilicity to ensure that low-entropy regions of the hydration shells forms at the binding sites. All the 50 protein quaternary structures were randomly selected from the protein data bank (PDB). According to our findings, shape-matching between the largest low-entropy hydration shell region of a protein subunit and that of the other subunit at binding sites is prevalent in all the investigated protein quaternary structures. This indicates that protein-protein binding is mainly guided by hydrophobic collapse between shape-matched low-entropy regions of hydration shells of individual proteins(*24, 45-47*). Space layout of the low-entropy region of the protein's hydration shell acts like a 'lock or key' for guiding the protein-protein binding in a precise manner.

As water molecules are ordered in low-entropy regions of hydration shells at the binding sites, they cause hydration shells in-between proteins to collapse hydrophobically, providing binding affinity. The shape-matching of the largest low-entropy hydration shell regions at the binding sites of the spike protein of the severe acute respiratory syndrome coronavirus (SARS-CoV-2) and the receptor angiotensin converting enzyme 2 (ACE2) is demonstrated in Fig.4. As seen at the RBD binding site, there are three isolated hydrophilic groups still exhibit their hydrophilicity, thereby disrupting the low-entropy structure of the hydration shell. In general, the fewer hydrophilic groups that can express their hydrophilicity at the protein's binding site, the higher the protein's binding affinity. The level of low-entropy nature of the hydration shell at a binding site can be evaluated by accounting the number of expressed hydrophilic groups at the binding site. We map the low-entropy hydration shell regions at the RBD binding sites of SARS-COV, SARS-COV-2 and 10 other SARS-COV-2 variants, respectively (see Fig.6). The proportions of the surface area of expressed hydrophilic groups on the low-entropy region at the binding sites of the RBDs are illustrated in Fig.7. The level of low-entropy nature of the hydration shell at the binding sites can be evaluated by the proportion of the surface area of expressed hydrophilic groups. The lower proportion means the lower level of entropy of the hydration shell at the binding sites. Namely, the lower level of low-entropy nature of the hydration shell at the binding site means the more binding affinity and the more transmission efficiency of the variant. The proportions of the surface area of expressed hydrophilic groups on the low-entropy region at the binding sites show that that magnitude on contagiousness these variants is Omicron BA.2, Omicron BA.1, Beta, Gamma, Mu, Alpha, Eta, Delta, Kappa, SARS-COV-2, Lambda, SARS-COV-1 in that order. Surprisingly, the results basically agree with the difference in contagiousness of the different variants deduced from epidemiological statistical evidence.

To further prove that the level of low-entropy nature of the hydration shell at the binding site is an important index reflecting the transmission efficiency of the different variants. The proportions of the surface area of expressed hydrophilic groups at the binding sites of six other viruses are counted in the data, as shown in Fig.8. The proportions of the surface area of expressed hydrophilic groups basically correctly reflect the basic reproduction number of these virus.

Experimental evidences had shown that mutations in the Kappa and Delta spike glycoproteins abrogate recognition by several monoclonal antibodies(48). The immune evasion by the Delta and Kappa SARS-CoV-2 variants may enable them spread in the countries with the highest vaccination rates. The Delta and Kappa SARS-CoV-2 variants share the L452R missense mutation in the RBD. The leucine (L) at amino-acid position 452 located at the center of a low-entropy region of the hydration shell, which is the binding sites for several antibodies. The change from leucine (L) to arginine (R) at amino-acid position 452 obviously decrease the low-entropy level of the hydration shell at the binding sites, due to the hydrophilicity expressed by the R (see Fig.9). This explain why the Kappa and Delta spike proteins can abrogate recognition by several monoclonal antibodies.

## 5. Conclusion

Hydrophilic groups at the binding site of a protein normally do not express their hydrophilicity to promote formation of a low-entropy hydration shell region at the binding site. Low-entropy regions of hydration shells around proteins drive hydrophobic attraction between shape-matched low-entropy regions of the hydration shells, which essentially coordinates protein-protein binding in rotational-conformational space of mutual orientations and determines their binding affinity. The bind affinity of a spike protein of a coronavirus to an acceptor protein of a host cell are resourced from long-range hydrophobic interaction force among the two low-entropy regions of the hydration shells at the two binding sites. Entropy increase caused by the hydrophobic collapse of the low-entropy hydration shells at the binding sites provides the binding affinity of protein docking. The spike-acceptor binding affinity is determined by the level of low-entropy of the hydration shell at the RBD binding sites of the spike protein. The level of low-entropy of the hydration shell at the binding site can be evaluated by calculating the proportion of the surface area of expressed hydrophilic groups at the binding site. Calculating the proportion of the surface area of expressed hydrophilic groups at binding sites can be used as a measure of the low-entropy level of the hydration shells at the binding sites. The low-entropy level of the hydration shell at the binding site is an important indicator of the contagiousness of the virus.

## Materials and Methods

## Protein structures

In this study, many experimentally determined native structures of proteins are used to study the protein-protein docking mechanism. All the three-dimensional (3D) structure data of protein molecules are resourced from the PDB database. IDs of these proteins according to PDB database are marked in all the figures. In order to show the distribution of low-entropy hydration shells on the surface of proteins at the binding sites in these figures, we used the structural biology visualization software PyMOL to display the low-entropy hydration shell areas. The mutations in the RBDs of the spikes of SARS-COV-2 variants are illustrated in the Supplementary.

## Hydrophilicity of residues

The detailed space layout of hydrophilicity of residues can be easily identified from the amount of charge of atoms according to the charmm36 force field (*49*).

## Acknowledgements

Lin Yang is indebted to Daniel Wagner from the Weizmann Institute of Science and Liyong Tong from the University of Sydney for their guidance. Lin Yang is grateful for his research experience in the Weizmann Institute of Science for inspiration. The authors acknowledge the financial support from the National Natural Science Foundation of China (Grant 21601054), Shenzhen Science and Technology Program (Grant No. KQTD2016112814303055), Natural Science Foundation of Heilongjiang Province (Grant LH2019F017), Science Foundation of the National Key Laboratory of Science and Technology on Advanced Composites in Special Environments, the Fundamental Research Funds for the Central Universities of China and the University Nursing Program for Young Scholars with Creative Talents in Heilongjiang Province of China (Grants UNPYSCT-2017126).

**Author Contributions** L.Yang, L.Ye, and X.H. formulated the study. L.Yang, J. L, S.G., X.M., C.H., H. Z., L.S., B.Z., and C.L. collected and analyzed the PDB data. C.H., L.Yang, C.L. wrote programs. L.Yang, L.Ye, Y.F. and X.H. wrote the paper, and all authors contributed to revising it. All authors discussed the results and theoretical interpretations.

## Additional Information

The authors declare no competing financial interests.

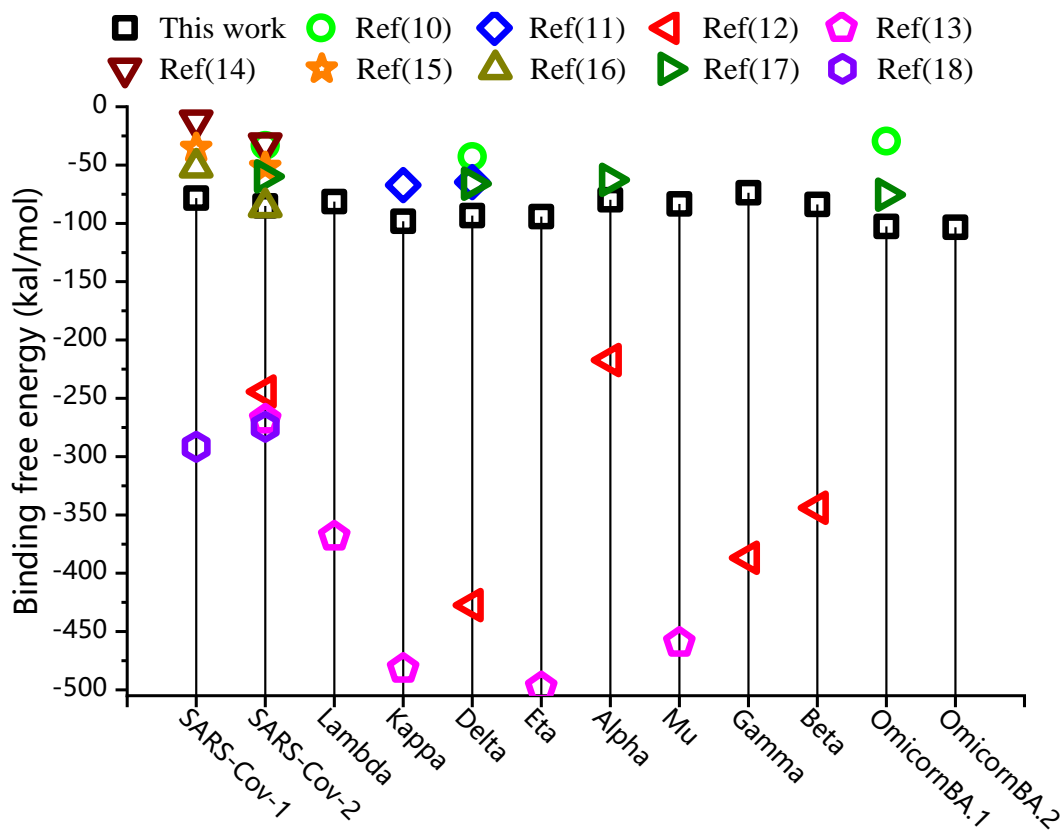

Fig.1 The RBD-ACE2 binding energies for SARS-CoV-2 variants were calculated by considering electrostatic interaction, hydrogen-bond interaction, van der Waals interaction, internal energy and nonpolar solvation energy

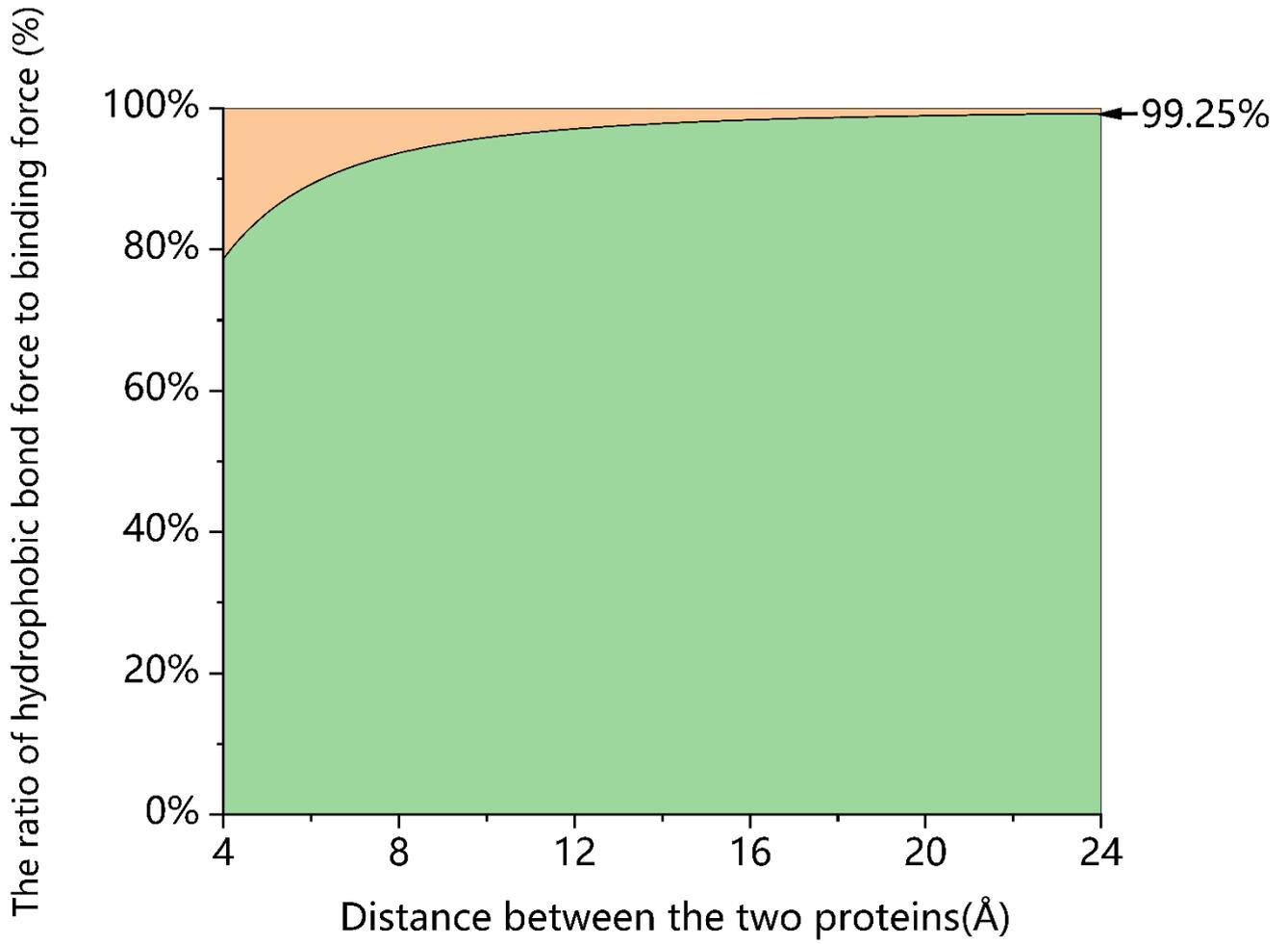

Fig.2 The ratio of hydrophobic bond force to the attractive force in-between the ACE2 and the SARS-COV-2 S RBD during the binding process.

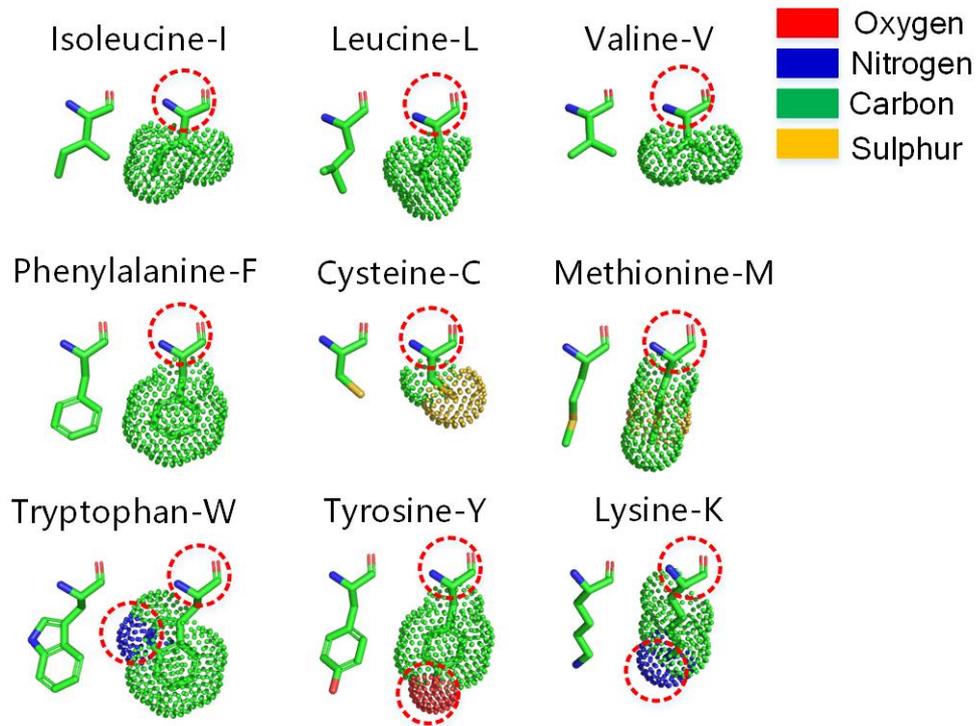

Fig.3 The residues with highly hydrophobic structures in the side-chains (hydrophobic portions are highlighted green and yellow). The hydrophilic groups can't express its hydrophilicity are marked by red dash circles.

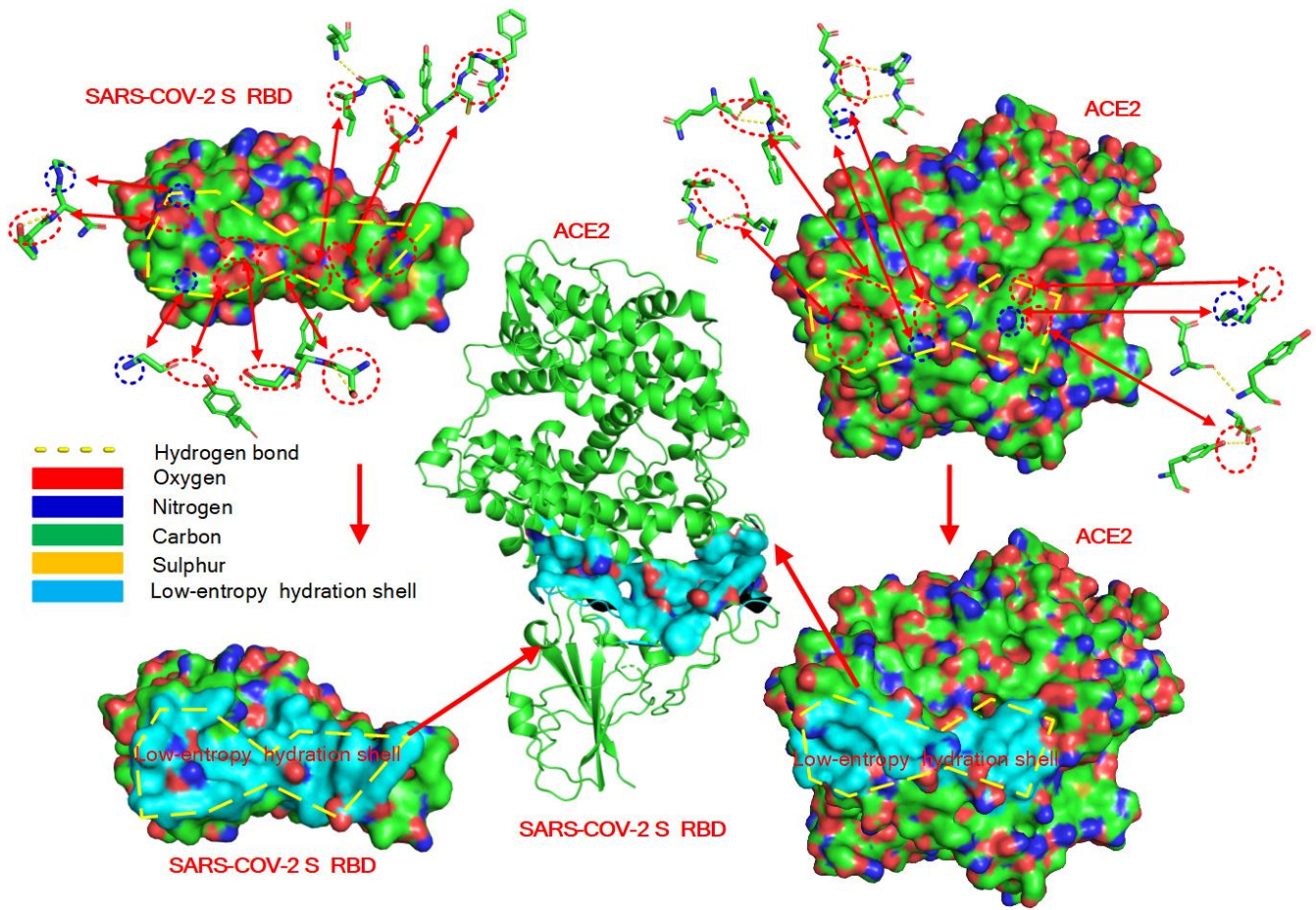

Fig. 4 Low-entropy hydration shells on the binding sites of the RBD of S protein of SARS-COV-2 and the ACE2. The binding sites of the two proteins are highlight by yellow dash lines, the low-entropy hydration shell region is highlighted in cyan color. The hydrophilic groups at the binding sites do not express their hydrophilicity are highlighted by red arrows.

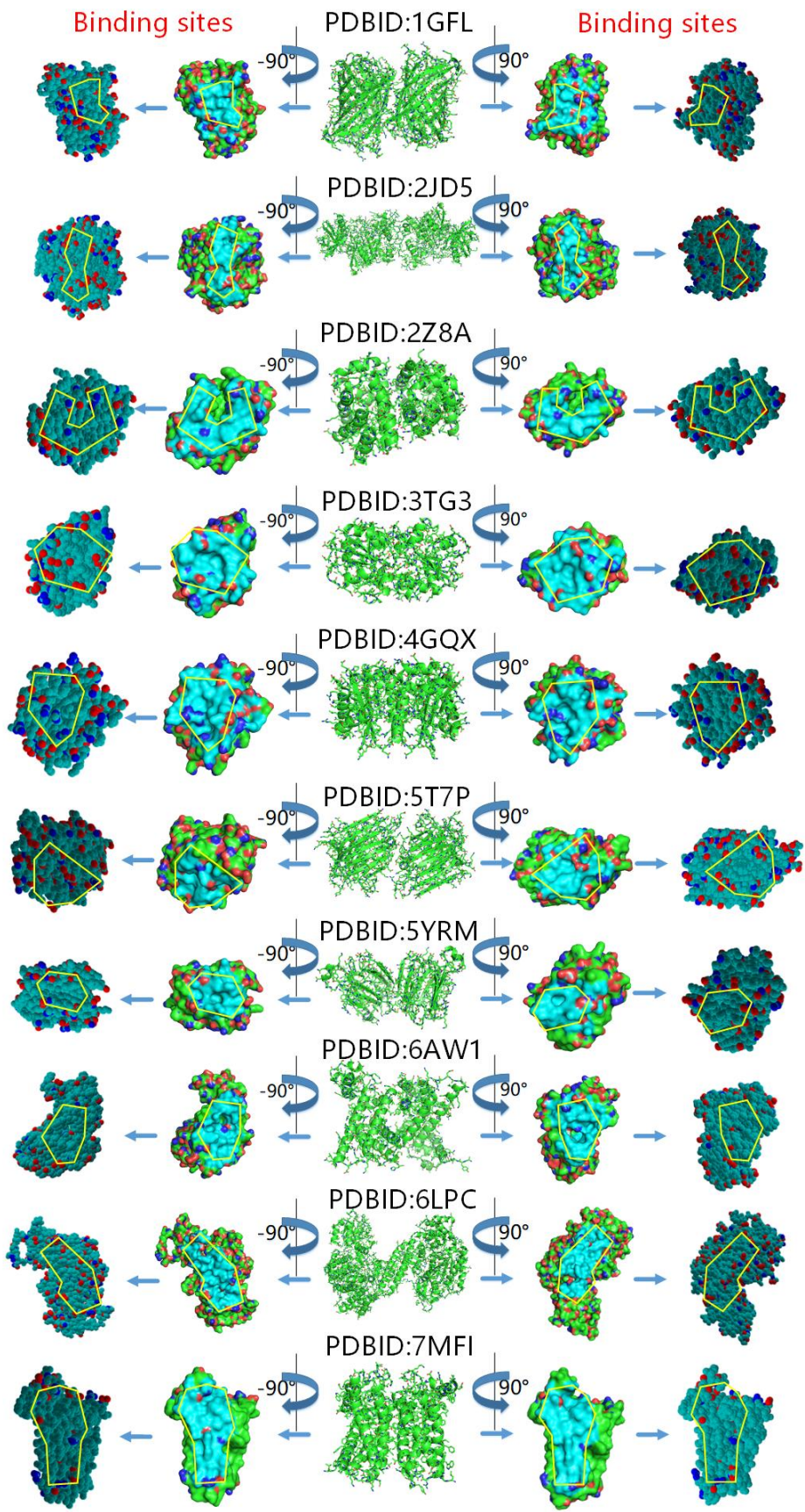

Fig.5 The prediction of binding sites of protein subunits for protein quaternary structures through identifying the largest low-entropy regions of hydration shells of individual protein subunits. The low-entropy hydration shell region is highlighted in cyan color.

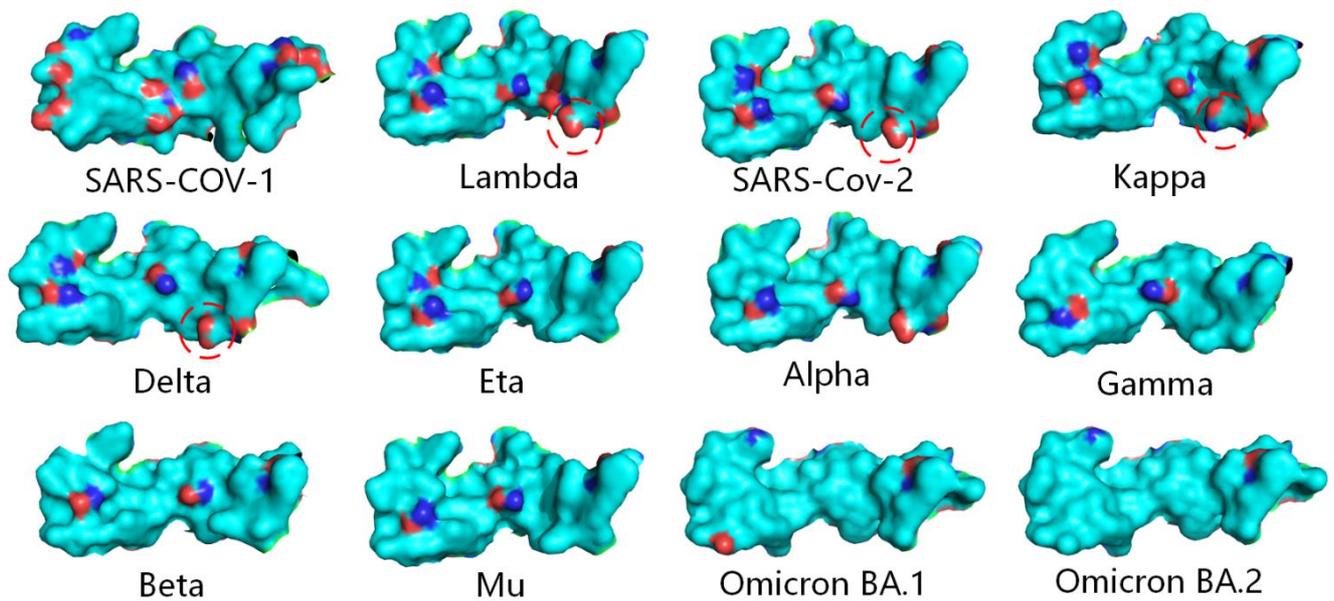

Fig.6 Distribution of the surface area of expressed hydrophilic groups on the low-entropy region at the binding sites of the RBDs of SARS-COV-1, SARS-COV-2, and SARS-COV-2 variants with ACE2. The low-entropy region of the hydration shells is highlighted in cyan color. The hydrophilic groups are not fully involved in the interfacial contact with the ACE2 partially are lighted by red dash circles. PDBIDs include: 6M17, 7R11, 7W9I, 7NXC, 7VX5, 7WBL, 2AJF.

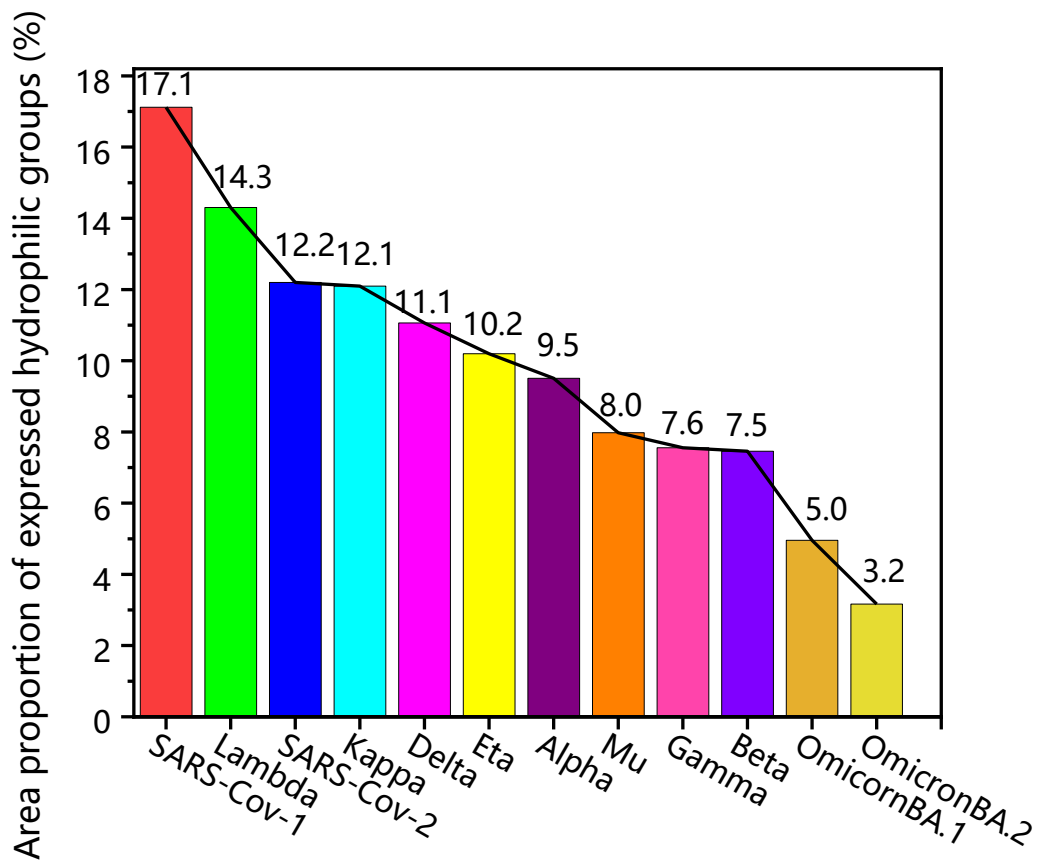

Fig.7 The proportion of the surface area of expressed hydrophilic groups on the low-entropy region at the binding sites of the RBDs of SARS-COV-1, SARS-COV-2, and SARS-COV-2 variants.

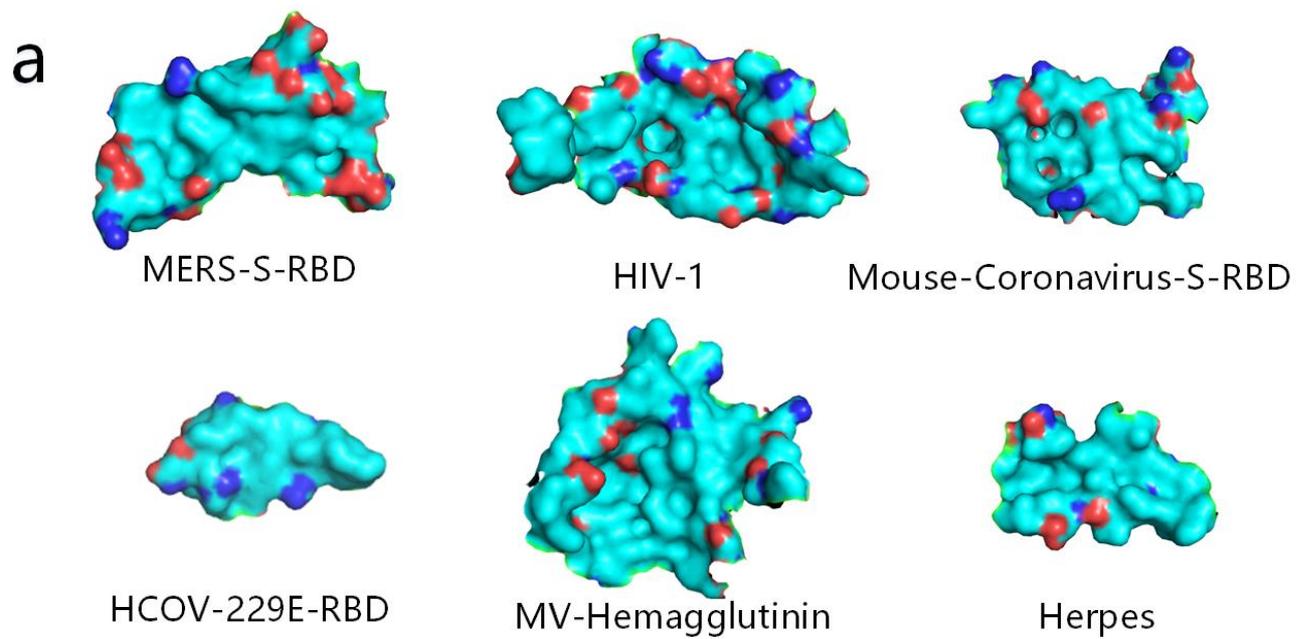

MERS-S-RBD     HIV-1     Mouse-Coronavirus-S-RBD

HCOV-229E-RBD     MV-Hemagglutinin     Herpes

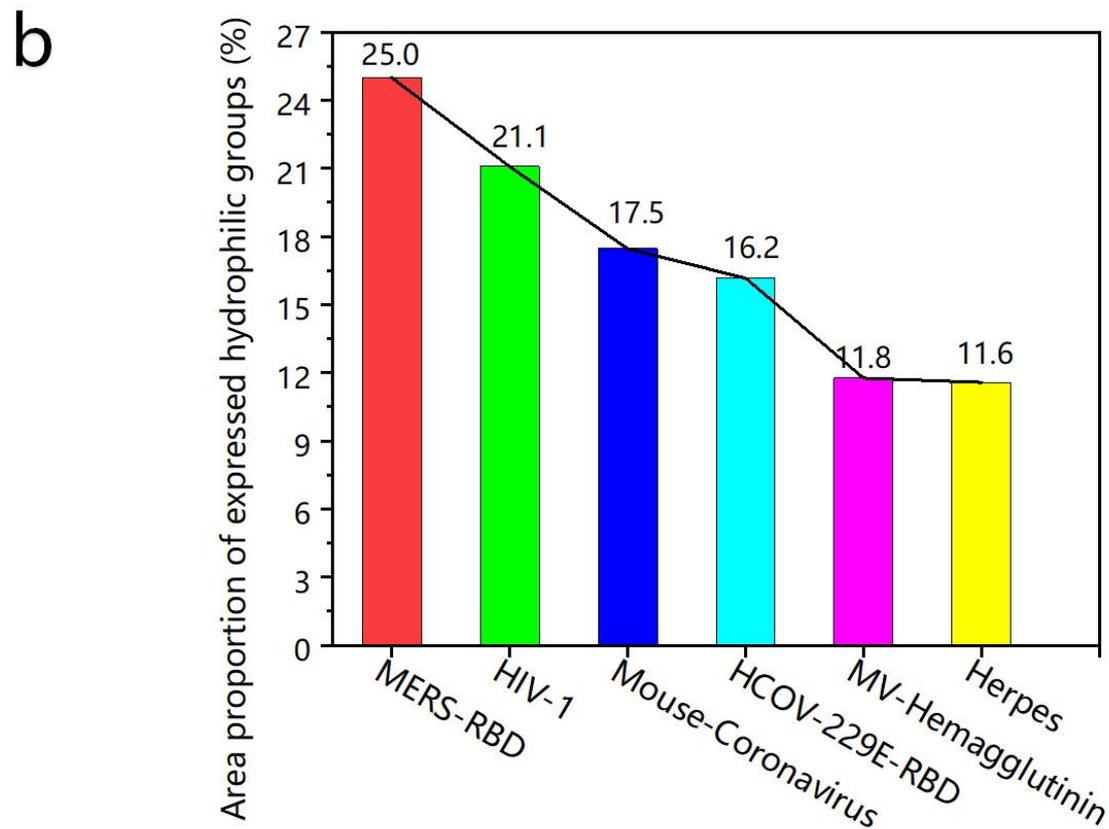

Fig.8 (a) The proportion of the surface area of expressed hydrophilic groups on the low-entropy region at the binding sites of MERS (PDBID: 4KR0), HIV (PDBID: 4RQS), Mouse-Coronavirus (PDBID: 6VSJ), human coronavirus HCoV-229E (PDBID: 6U7G) and MV-Hemagglutinin (PDBID: 3ALZ), herpes. (PDBID: 1JMA). The low-entropy regions of hydration shell are highlighted in cyan color. (b) The proportion of the surface area of expressed hydrophilic groups on the low-entropy region at the binding sites.

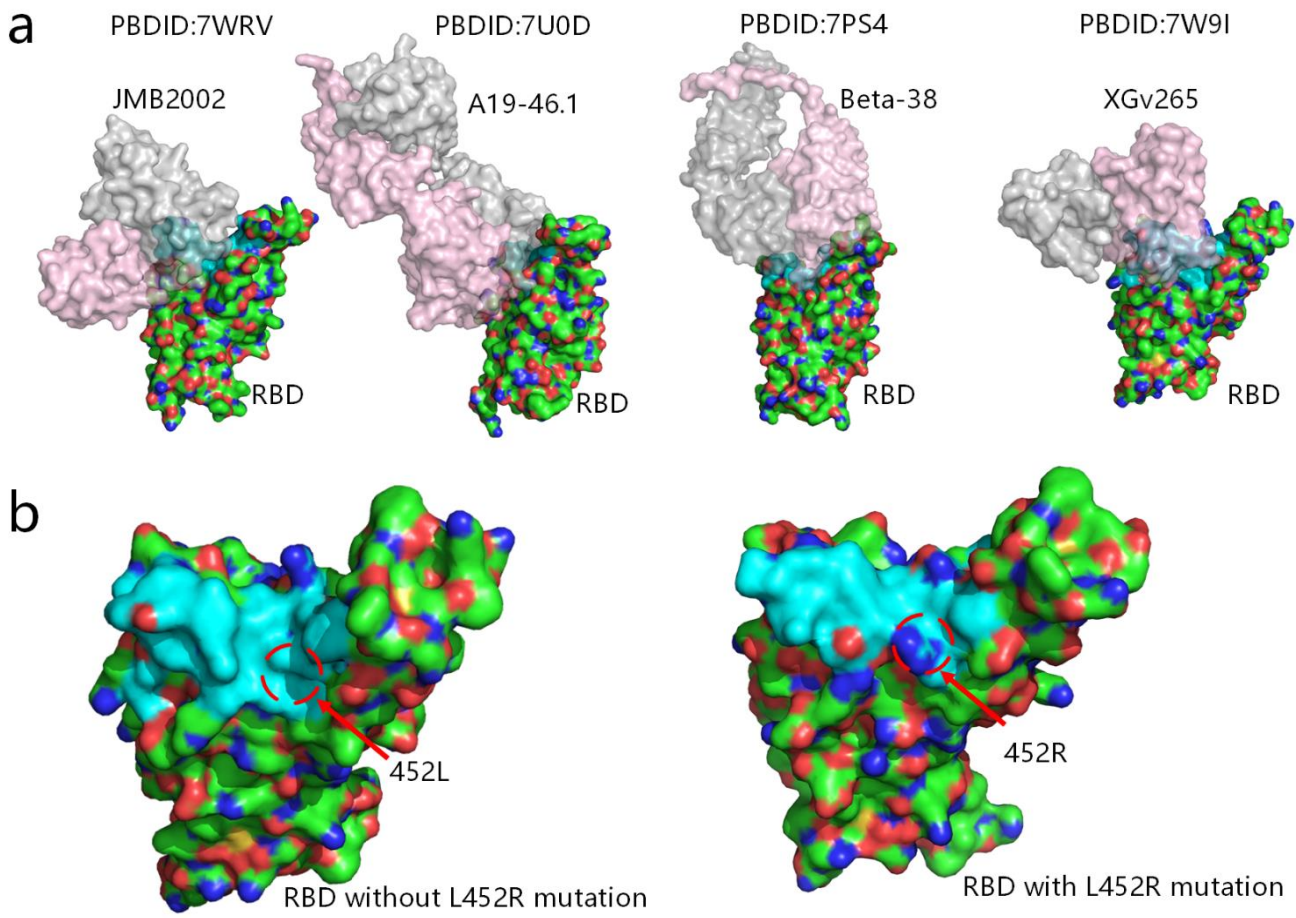

Fig.9 (a) The antibodies bind with the spike protein of SARS-COV-2 covering the at amino-acid position 452. (b) The low-entropy region at amino-acid position 452 of the RBD of SARS-COV-2 before and after the mutation of L452R.

# Supplementary

**Molecular dynamics simulations of the binding free energy**

To assess the binding free energy of the SARS-CoV Spike-RBD/ACE2 complex, we performed all-atom MD simulations of Wild-type (WT) and variants. The three-dimensional structure of the Spike protein RBD associated with ACE2 was retrieved from Protein Data Bank(1). Protein complexes were considered in this work, namely SARS-CoV-1 Spike-RBD/ACE2 (PDB ID: 2AJF)(2), SARS-CoV-2 Spike-RBD/ACE2 (PDB ID: 6M0J)(3), Alpha (B.1.1.7)、Beta (B.1.351) Spike-RBD/ACE2 (PDB ID: 7R10、7R11)(4), Delta (B.1.617.2) Spike-RBD/ACE2 (PDB ID: 7W9I)(5), Gamma (P.1) Spike-RBD/ACE2 (PDB ID: 7NXC)(6), Kappa Variant Spike-RBD/ACE2 (PDB ID: 7VX5)(7), Omicron BA.1 Spike-RBD/ACE2 (PDB ID: 7R10)(8). The variants namely, Eta (B.1.525), Lambda (C.37), Mu (B.1.621) and Omicron) (BA.2) were generated by mutating specific residues in Wild-type Spike RBD using Pymol(9).

All atomistic MD simulation was carried out using Gromacs 2021.1 software(10) using Amber ff14SB force field parameters(11) and TIP3P water model(12). A cubic simulation box consisting of 0.15mM NaCl was created for the systems under periodic boundary conditions (PBCs). The distance from the protein solute complex to the wall of the cube was set to 10 Å. All bonds and heavy atoms were restricted by the LINCS (13) constraint algorithms allowed for an integration timestep of 2 fs. . The cutoff values of short-range nonbonded electrostatic and Lennard-Jones interactions were set to 12 Å. Particle-mesh Ewald method (14) was used to treat the long-range electrostatic interactions with a 1.2 Å grid spacing.

To remove bad contacts from the initial structure which might have been created due to the new mutations and addition of water and ions,energy minimization was subjected to in Steepest Descent algorithm. The systems were equilibrated at 300 K in NVT ensemble (constant number of particles, volume, and temperature) using V-rescale coupling algorithm (15) for about 500ps and then equilibrated in NPT ensemble(constant number of particles, pressure, and temperature) using 1 atmospheric pressure using Parrinello-Rahman barostat(16) for 500ps. The ensemble with 1000 kJ·mol⁻¹·nm⁻² harmonic constraints on the heavy atoms of the proteins aims to equilibrate water and ions. After the potential energy and density of the system have fully converged, the atomic restrictions were removed and production simulation was carried out for the 12 protein complexes for 50 ns in NPT . VMD were used for visualization of the trajectories (17).

**Binding energy calculation between RBD and ACE2**

The binding free energy between RBD and ACE2 for WT and variants were computed by using the Molecular Mechanics/Poisson Boltzmann Surface Area (MM/PBSA) employed in the gmx_MMPBSA tool(18, 19), which is an end-point method. In this methodology, the binding free energy ($\Delta G$bind) between the proteins is calculated by

$$\Delta G_{\text{bind}} = \Delta G_{\text{complex}} - \Delta G_{\text{recepter}} - \Delta G_{\text{ligand}}$$

$$\Delta G_{\text{bind}} = \Delta H - T\Delta S \approx \Delta E_{\text{MM}} + \Delta G_{\text{solv}} - T\Delta S$$

$$\Delta E_{\text{MM}} = \Delta E_{\text{bond}} + \Delta E_{\text{angle}} + \Delta E_{\text{dihedral}} + \Delta E_{\text{vdw}} + \Delta E_{\text{coulomb}}$$

$$\Delta G_{\text{solv}} = \Delta G_{\text{polar}} + \Delta G_{\text{nonpolar}}$$

where $\Delta G_{\text{complex}}$, $\Delta G_{\text{recepter}}$, and $\Delta G_{\text{ligand}}$ represent the total free energies of the complex, the receptor, and the ligand, respectively. Further, $\Delta G$bind can be usually decomposed into three terms: the average molecular mechanic's potential energy in the vacuum $\Delta E_{\text{MM}}$, the free energy of solvation $\Delta G$sol, and the conformational entropy $-T\,\Delta S$, here $S$ and T denote the entropy and temperature, respectively. The $\Delta E_{\text{MM}}$ consists of bonded terms which include the bond stretching $\Delta E_{\text{bond}}$, angle bending $\Delta E_{\text{angle}}$ and dihedral angles $\Delta E_{\text{dihedral}}$, and nonbonded terms, which include the electrostatic $\Delta E_{coulomb}$ and the Van der waal interactions $\Delta E_{vdw}$. The solvation free energy, $\Delta G_{solv}$ takes both electrostatic and non-electrostatic ($\Delta G_{polar}$ and $\Delta G_{nonpolar}$) components. The vacuum electrostatic dielectric constant and the solvent dielectric constant were set to 2 and 80, respectively. The binding free energies for the complexes were calculated by taking 1000 snapshots at 30-ps intervals in the last 30 ns.

*Corresponding author. E-mail address: linyang@hit.edu.cn (Lin Yang) Hexd@hit.edu.cn (Xiaodong He) ¹These authors contributed equally to this work.

**Examples of the prediction of binding sites of protein subunits**

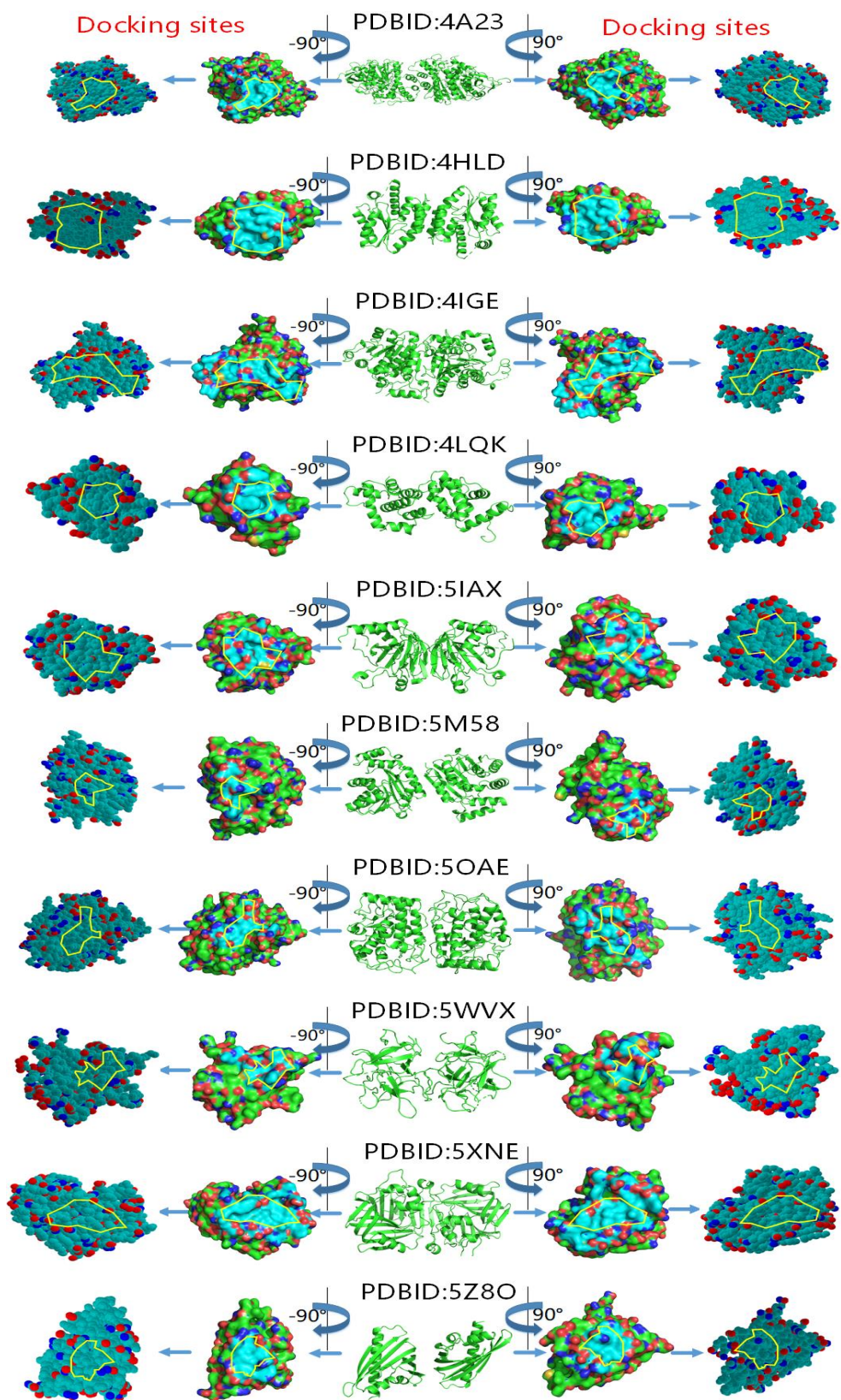

Fig.1 The prediction of binding sites of protein subunits for 10 protein quaternary structures through identifying the largest low-entropy regions of hydration shells of individual protein subunits.

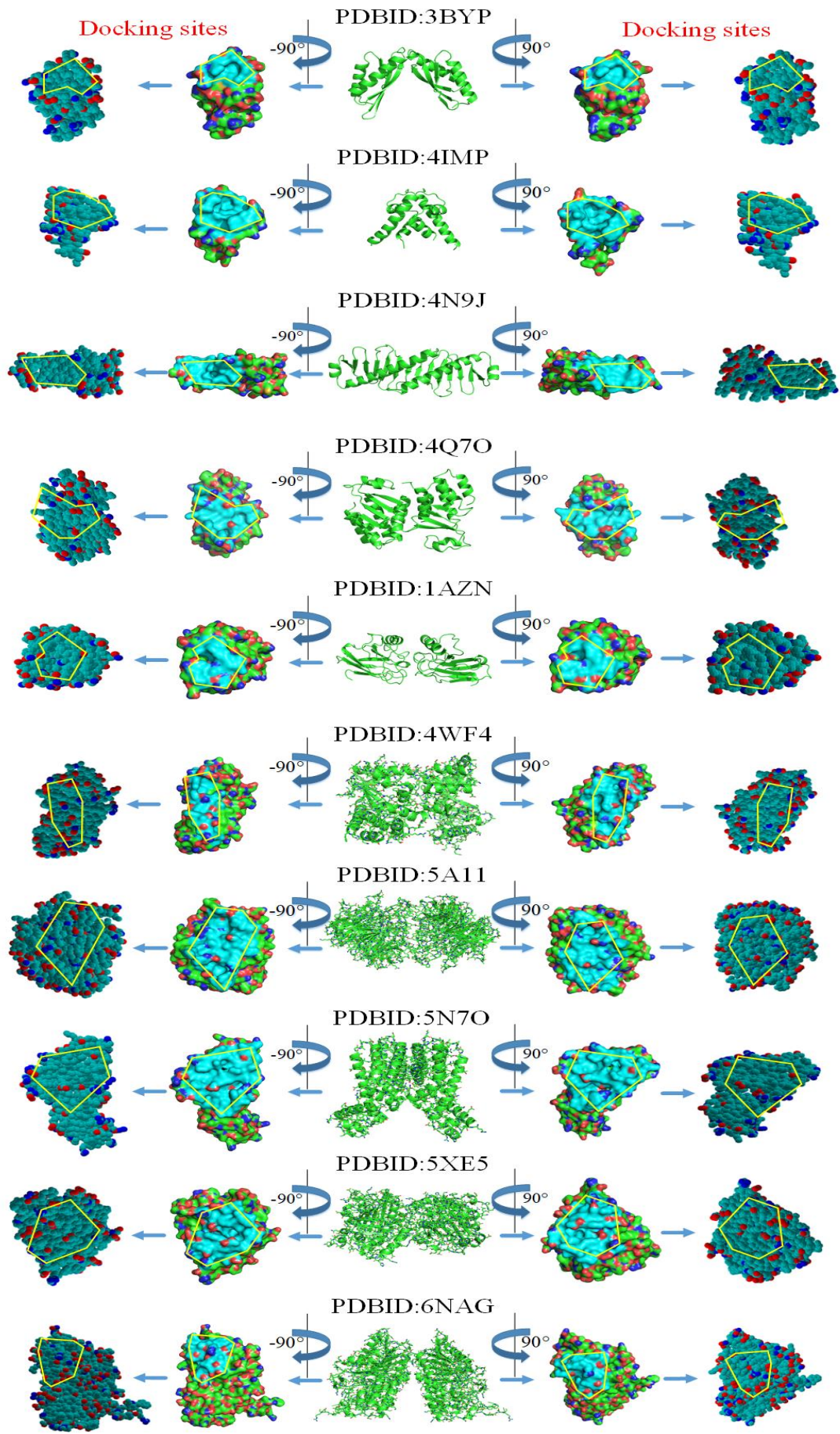

Fig.2 The prediction of binding sites of protein subunits for 10 protein quaternary structures through identifying the largest low-entropy regions of hydration shells of individual protein subunits.

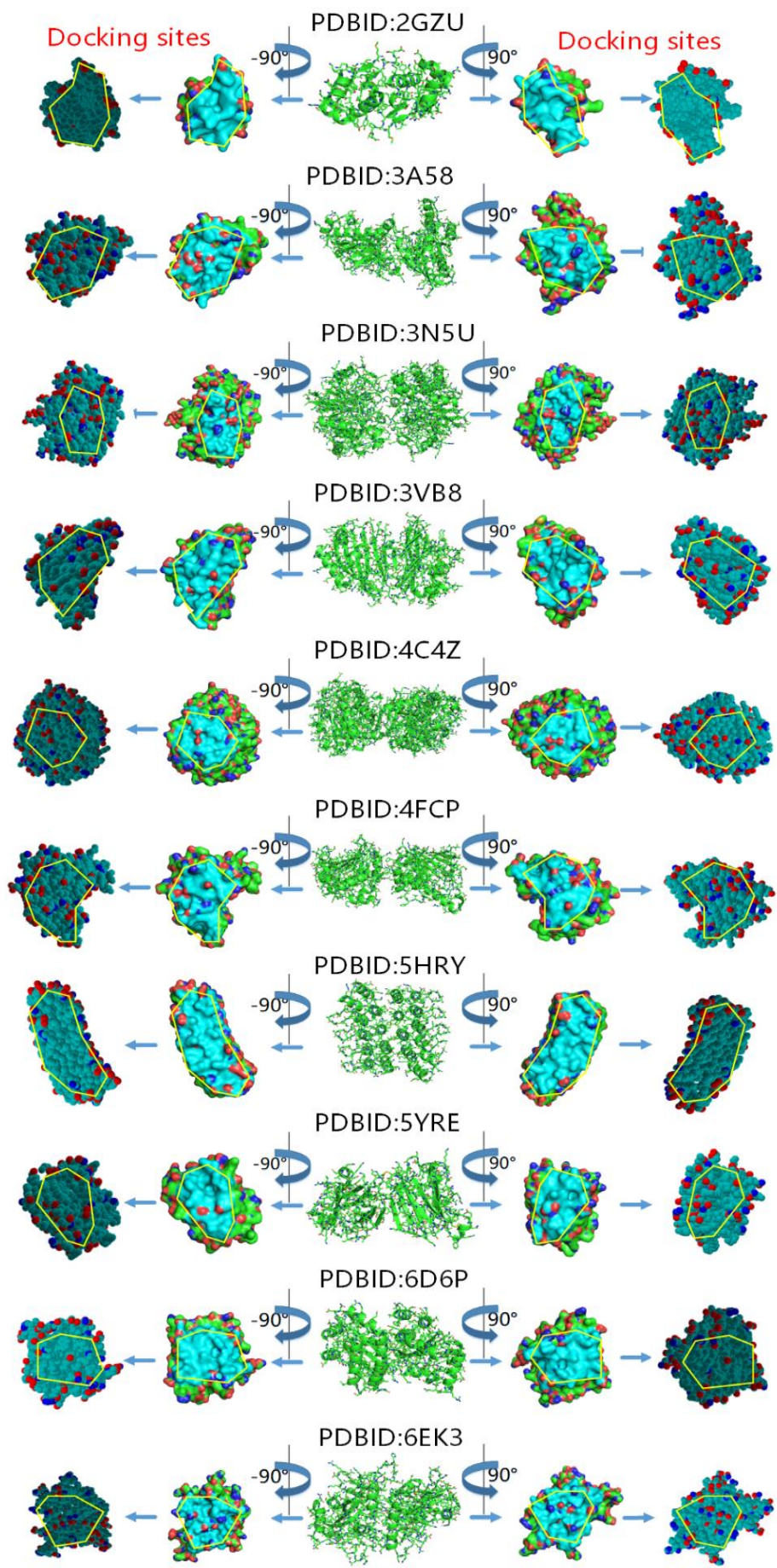

Fig.3 The prediction of binding sites of protein subunits for 10 protein quaternary structures through identifying the largest low-entropy regions of hydration shells of individual protein subunits.

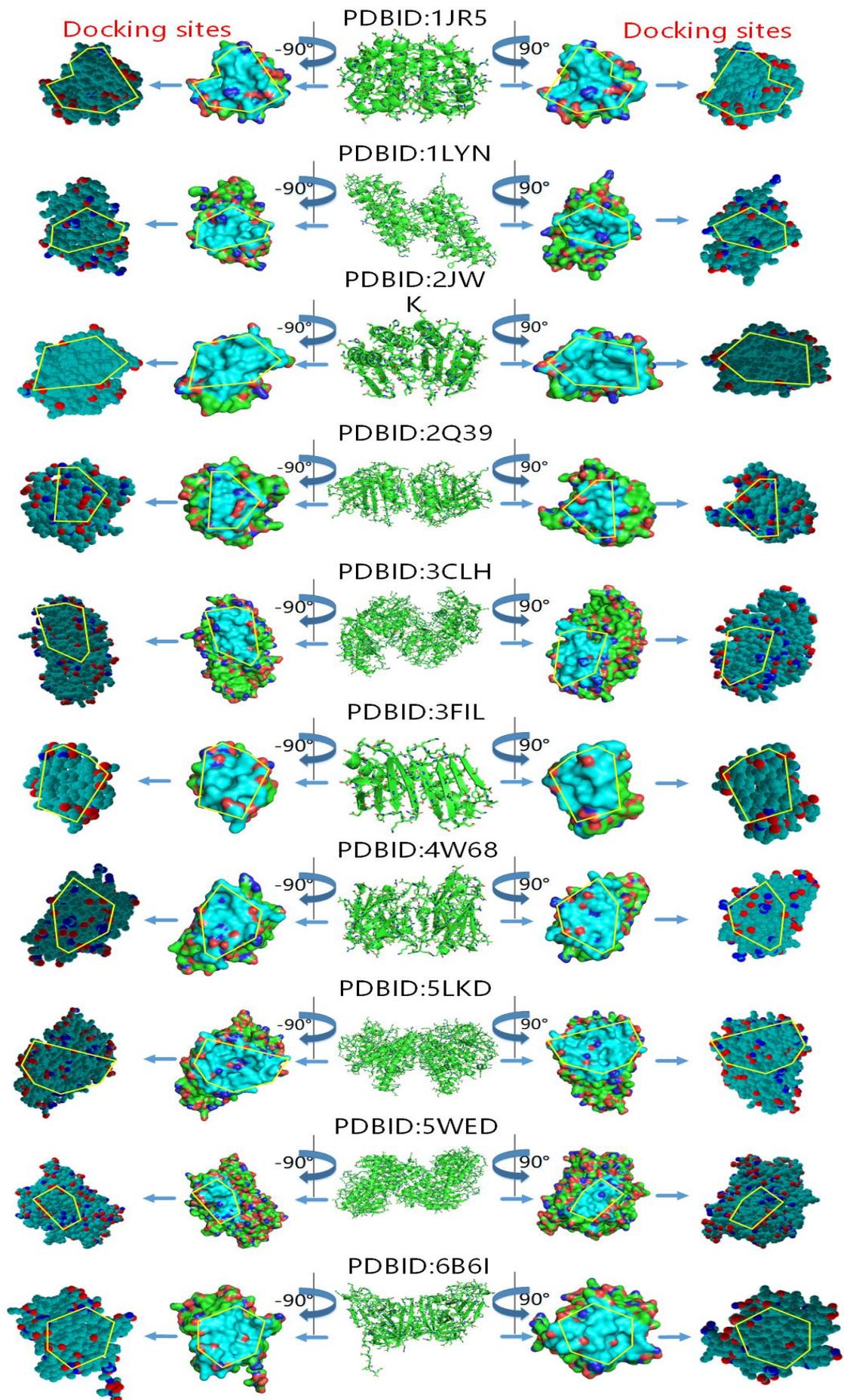

Fig.4 The prediction of binding sites of protein subunits for 10 protein quaternary structures through identifying the largest low-entropy regions of hydration shells of individual protein subunits.

# The mutations in the RBDs of the spikes of SARS-COV-2 variants

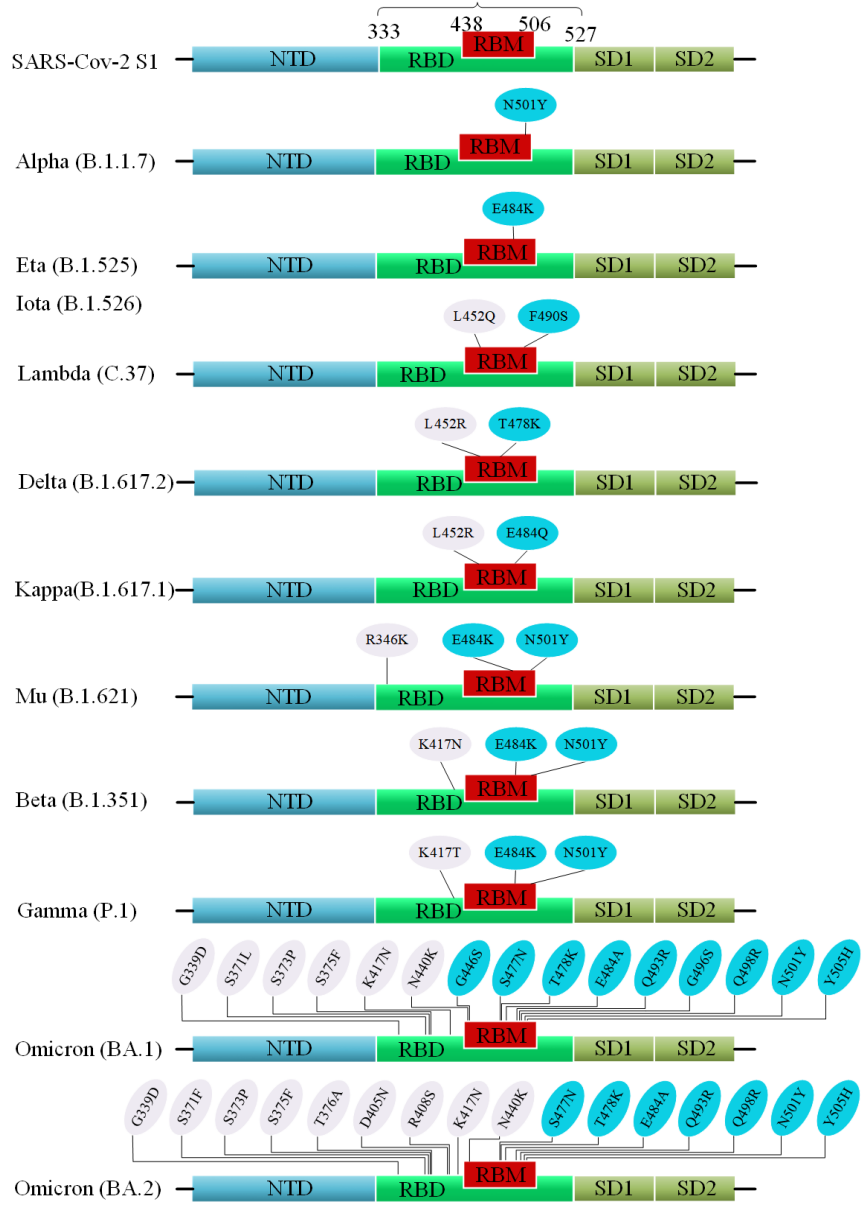

Fig.5 The mutations in the RBDs of the spikes of SARS-COV-2 variants.